\overfullrule=0pt
\input harvmac

\lref\Bnon{
  N.~Berkovits,
  ``Pure spinor formalism as an N=2 topological string,''
JHEP {\bf 0510}, 089 (2005).
[hep-th/0509120].
%%CITATION = hep-th/0509120%%
}

\lref\Bpure{
  N.~Berkovits,
  ``Super Poincare covariant quantization of the superstring,''
JHEP {\bf 0004}, 018 (2000).
[hep-th/0001035].
%%CITATION = hep-th/0001035%%
}

\lref\explaining {
N.~Berkovits,
``Explaining the Pure Spinor Formalism for the Superstring,''
JHEP {\bf 0801}, 065 (2008).
[arXiv:0712.0324 [hep-th]].
%%CITATION = arXiv:0712.0324%%
}

\lref\osv{
  O.~Chandia,
  ``The b Ghost of the Pure Spinor Formalism is Nilpotent,''
Phys.\ Lett.\ B {\bf 695}, 312 (2011).
[arXiv:1008.1778 [hep-th]].
%%CITATION = arXiv:1008.1778%%
}

\lref\rennan{
  R.~L.~Jusinskas,
  ``Nilpotency of the b ghost in the non minimal pure spinor formalism,''
[arXiv:1303.3966 [hep-th]].
%%CITATION = arXiv:1303.3966%%
}

\lref\BeisertJR{
  N.~Beisert, C.~Ahn, L.~F.~Alday, Z.~Bajnok, J.~M.~Drummond, L.~Freyhult, N.~Gromov and R.~A.~Janik {\it et al.},
  ``Review of AdS/CFT Integrability: An Overview,''
Lett.\ Math.\ Phys.\  {\bf 99}, 3 (2012).
[arXiv:1012.3982 [hep-th]].
%%CITATION = arXiv:1012.3982%%
}

%\SiegelYD
\lref\SiegelYD{
  W.~Siegel,
  %``AdS/CFT in superspace,''
[arXiv:1005.2317 [hep-th]].
%%CITATION = arXiv:1005.2317%%
}

\def\bar{\overline}

\def\a{{\alpha}}

\def\l{{\lambda}}
\def\lb{{\overline\lambda}}

\def\lb{{\overline\lambda}}
\def\b{{\beta}}

\def\g{{\gamma}}
\def\G{{\Gamma}}
\def\Gb{{\bar\Gamma}}

\def\d{{\delta}}

\def\half{{1\over 2}}
\def\p{{\partial}}

\def\pb{{\overline\partial}}
\def\t{{\theta}}
\def\tt{{\tilde\theta}}

\def\lb{{\bar{\lambda}}}

\def\mn{{ {{(\l\g_m\g_n\lb)}\over{2(\l\lb)}} }}

\Title{\vbox{\baselineskip12pt
\hbox{ICTP-SAIFR/2013-006 }}}
{{\vbox{\centerline{Dynamical Twisting and the $b$ Ghost}
\smallskip
\centerline{in the Pure Spinor Formalism}}} }
\bigskip\centerline{Nathan Berkovits\foot{e-mail: nberkovi@ift.unesp.br}}
\bigskip
\centerline{\it ICTP South American Institute for Fundamental Research}
\centerline{\it Instituto de F\'\i sica Te\'orica, UNESP - Univ. 
Estadual Paulista }
\centerline{\it Rua Dr. Bento T. Ferraz 271, 01140-070, S\~ao Paulo, SP, Brasil}
\bigskip

\vskip .3in

After adding an RNS-like fermionic vector $\psi^m$ to the pure spinor formalism,
the non-minimal $b$ ghost takes a simple form similar to the pure spinor
BRST operator. The N=2 superconformal field theory generated by the $b$ ghost
and the BRST current can be interpreted as a ``dynamical twisting'' of the
RNS formalism where the choice of which
spin $\half$ $\psi^m$ variables are twisted into spin 0
and spin 1 variables is determined by the pure spinor variables that 
parameterize the coset $SO(10)/U(5)$.

\vskip .3in

\Date {May 2013}

\newsec{Introduction}

The pure spinor 
formalism for the superstring \Bpure\ has the advantage over the 
Ramond-Neveu-Schwarz (RNS)
formalism of being manifestly
spacetime supersymmetric and has the advantage over the
Green-Schwarz (GS) formalism of allowing covariant quantization. However, the
worldsheet origin of the pure spinor formalism is mysterious since its BRST
operator and $b$ ghost do not arise in an obvious manner from gauge-fixing.

In the non-minimal pure spinor formalism,
the BRST current and $b$ ghost 
can be interpreted as twisted $\hat c=3$
N=2 superconformal generators \Bnon. But when expressed in terms of the d=10
superspace variables and the non-minimal pure spinor variables,
the $b$ ghost and the resulting N=2 superconformal transformations are 
extremely complicated. In fact, the nilpotency of the $b$ ghost was only
recently verified \osv\rennan.

In this paper, it will be shown that the $b$ ghost dramatically simplifies
when expressed in terms of a fermionic vector $\psi^m$ that is defined in
terms of the other worldsheet variables. 
If one treats the ten $\psi^m$ variables as independent variables,
5 of the 16 $\t^\a$ variables of d=10 superspace 
(and their conjugate momenta) can
be eliminated \explaining. 
The 
remaining 11 $\t^\a$ variables and their conjugate momenta
transform as the worldsheet superpartners of the pure spinor variables.
The resulting N=2 superconformal field theory generated by the $b$ ghost and
the BRST current can be interpreted as a 
``dynamically twisted'' version of the RNS formalism.

In this dynamically twisted superconformal field theory, the 
N=2 generators are
\eqn\introd{ T =- \half \p x^m \p x_m - 
{{(\l\g_m\g_n\lb)}\over{2(\l\lb)}} \psi^m \p\psi^n + ...,}
$$b = \mn \psi^m \p x^n + ..., $$
$$j_{BRST} =- \mn \psi^n\p x^m + ...,$$
$$J = -\mn \psi^m \psi^n + ...,$$
where $\l^\a$ and $\lb_\a$ are the non-minimal pure spinor ghosts whose 
projective components
parameterize the coset ${{SO(10)}}/{{U(5)}}$ that describes different twistings.
The remaining terms $...$ in \introd\ are determined by requiring that
$(\l^\a,\lb_\a)$ and their worldsheet superpartners transform in an N=2
supersymmetric manner.

So the resulting N=2 superconformal field theory is the sum of a dynamically
twisted RNS
superconformal field theory with an N=2 superconformal field theory for the pure
spinor variables. 
This interpretation of the BRST operator and the $b$ ghost as coming from
dynamical twisting of an N=1 superconformal field theory
will hopefully lead to a better geometrical understanding of
the pure spinor formalism.

In section 2, the non-minimal pure spinor formalism is reviewed. In section 3,
the $b$ ghost in the pure spinor formalism
is shown to simplify when expressed in terms of an
RNS-like $\psi^m$ variable. In section 4, dynamical twisting of
the RNS formalism will be defined and the
resulting twisted N=2 superconformal generators
will be
related to the $b$ ghost and
BRST current in the pure spinor formalism.
And in section 5, the results will be summarized.

\newsec{Review of Non-Minimal Pure Spinor Formalism}

As discussed in \Bnon,
the left-moving contribution to the worldsheet
action in the non-minimal pure spinor formalism
is
\eqn\nmps{S = \int d^2 z [-{1\over 2 }\p x^m \pb x_m - p_\a \pb \t^\a +
w_\a \pb \l^\a +\bar w^\a \pb \lb_\a - s^\a \pb r_\a]}
where $x^m$ and $\t^\a$ are d=10 superspace variables for $m=0$ to 9 and
$\a=1$ to 16, $p_\a$ is the conjugate
momentum to $\t^\a$, 
$\l^\a$ and $\lb_\a$ are bosonic Weyl and anti-Weyl pure spinors constrained
to satisfy $\l\g^m\l=0$ and $\lb\g^m\lb=0$, and $r_\a$ is a fermionic spinor
constrained to satisfy $\lb\g^m r=0$. Because of the constraints on the
pure spinor variables,
their conjugate momenta
$w_\a$, $\bar w^\a$ and $s^\a$ can only appear in gauge-invariant
combinations such as
\eqn\combi{N^{mn} = \half (w\g^{mn}\l), \quad J_\l = (w\l),\quad
S^{mn} = \half(s\g^{mn} \lb), \quad S = (s \lb),}
which commute with the pure spinor constraints.

The d=10 superspace variables satisfy the free-field OPE's
\eqn\opes{ x^m(y) x^n(z) \to - \eta^{mn} \log |y-z|^2, \quad 
p_\a (y) \t^\b(z) \to (y-z)^{-1}\d_\a^\b,}
and,
as long as the pure spinor
conjugate momenta appear in gauge-invariant combinations and normal-ordering
contributions are ignored,
one can use the free-field OPE's of pure spinor variables 
\eqn\opep{ w_\a(y) \l^\b(z) \to (y-z)^{-1} \d_\a^\b, \quad
\bar w^\a(y) \lb_\b(z) \to (y-z)^{-1} \d^\a_\b, \quad
s^\a(y) r_\b(z) \to (y-z)^{-1} \d^\a_\b.}
It is convenient to define the spacetime 
supersymmetric combinations
\eqn\susy{\Pi^m = \p x^m+\half(\t\g^m\p\t), \quad 
d_\a = p_\a -\half (\p x^m + {1\over 4}
(\t\g^m\p\t)) (\g_m\t)_\a}
which satisfy the OPE's
\eqn\opesuper{d_\a(y) d_\b(z) \to - (y-z)^{-1} \Pi_m \g^m_{\a\b}, \quad
d_\a (y) \Pi^m(z) \to (y-z)^{-1} (\g^m\p\t)_\a.}

As shown in \Bnon, the non-minimal BRST current 
forms a twisted $\hat c=3$ N=2 superconformal
algebra with the stress tensor, a composite $b$ ghost, and a U(1) ghost-number
current. These twisted N=2 generators are
\eqn\purestress{ T =  -\half \p x^m \p x_m - p_\a \p\t^\a + w_\a \p\l^\a
+\bar w^\a \p\lb_\a - s^\a \p r_\a,}
\eqn\nonminb{b =  s^\a\p\lb_\a  + {{\lb_\a (2
\Pi^m (\g_m d)^\a-  N_{mn}(\g^{mn}\p\t)^\a
- J_\l \p\t^\a -{1\over 4} \p^2\t^\a)}\over{4(\lb\l)}} }
$$- {{(\lb\g^{mnp} r)(d\g_{mnp} d +24 N_{mn}\Pi_p)}\over{192(\lb\l)^2}}
+ {{(r\g_{mnp} r)(\lb\g^m d)N^{np}}\over{16(\lb\l)^3}} -
{{(r\g_{mnp} r)(\lb\g^{pqr} r) N^{mn} N_{qr}}\over{128(\lb\l)^4}}, $$
\eqn\nonminbrst{ j_{BRST} = \l^\a d_\a - \bar w^\a r_\a,}
\eqn\nonminu{ J_{ghost} = w_\a \l^\a - s^\a r_\a - 2(\l\lb)^{-1}
[(\l\p\lb) + (r\p\t)] + 2(\l\lb)^{-2}(\l r)(\lb\p\t).}

The terms $-{1\over {16}}(\l\lb)^{-1}\p^2 \t^\a$ in \nonminb\ and 
$-2(\l\lb)^{-1}
[(\l\p\lb) + (r\p\t)] + 2(\l\lb)^{-2}(\l r)(\lb\p\t)$ in 
\nonminu\ are higher-order in $\a'$ and
come from normal-ordering contributions. To simplify the analysis, these
normal-ordering contributions 
will be ignored throughout this paper. However, it
should be possible to do a more careful analysis which takes into
account these contributions.

\newsec{Simplification of $b$ Ghost}

In this section, the complicated expression of \nonminb\ for the $b$ ghost 
will be simplified by including an auxiliary fermionic vector variable
which will be later related to the RNS $\psi^m$ variable.
The trick to simplifying the $b$ ghost is to observe that the terms
involving $d_\a$ in \nonminb\ always appear in the combination
\eqn\Gc{
\bar\G^m = \half(\l\lb)^{-1} (\lb\g^m d) -{1\over 8}
 (\l\lb)^{-2} (\lb\g^{mnp}r) N_{np}. }
Note that only five components of $\bar\G^m$ are independent since
$\bar\G^m (\g_m\lb)^\a =0$.
In terms of $\Gb^m$,
\eqn\bg
{b = \Pi^m \Gb_m -{1\over 4} (\l\lb)^{-1}(\l\g^{mn}r)\Gb_m\Gb_n +
s^\a\p\lb_\a + w_\a\p\t^\a -\half (\l\lb)^{-1}(w\g_m\lb) (\l\g^m\p\t) }
where terms coming from normal-ordering are being ignored and the identity
\eqn\ident{\d_\b^\g \d_\a^\d = \half \g^m_{\a\b}\g_m^{\g\d}
-{1\over 8} (\g^{mn})^\g_\a (\g_{mn})^\d_\b -{1\over 4}\d_\a^\g \d_\b^\d}
has been used.

It is useful to treat \Gc\ as a first-class constraint where $\Gb^m$
is a new worldsheet variable
which carries $+1$ conformal weight and satisfies the constraint
$\Gb^m (\g_m\lb)^\a=0$. Its conjugate momentum will be defined as $\G_m$
of conformal weight zero and can only appear in combinations invariant under
the gauge tranformation generated by the constraint of \Gc.
Note that
$\Gb^m$ and $\G_m$ satisfy the OPE $\Gb^m(y) ~\G^n(z) \to (y-z)^{-1}
\eta^{mn}$ and have no singular OPE's with the other variables.

One can easily verify that the $b$ ghost of \bg\ is gauge-invariant
since it has no singularity with
\Gc. Furthermore, any operator ${\cal O}$ 
which is independent of $\G_m$ can be written
in a gauge-invariant manner by defining
${\cal O}_{inv} = e^R ~{\cal O} ~ e^{-R}$ where
\eqn\gaugeR{R = \int \G_m
[\half(\l\lb)^{-1} (\lb\g^m d) -{1\over 8} 
(\l\lb)^{-2} (\lb\g^{mnp}r) N_{np}]. }

For example, the gauge-invariant version of the BRST current is 
\eqn\BRSTg{G^+ = e^R~ (\l^\a d_\a - \bar w^\a r_\a ) e^{-R} =
\l^\a d_\a - \bar w^\a r_\a}
$$- \half\G^m (\l\lb)^{-1} 
[(\lb \g_m \g_n \l) \Pi^n - (r \g_n \g_m \l) \Gb^n] $$
$$
+{1\over 4}\G^m\G^n [ (\l\lb)^{-1}(\lb \g_{mn}\p\t) - 
(\l\lb)^{-2}(\lb\p\t)(\lb\g_{mn}\l)]$$
$$ + {1\over 8}\G^m\G^n (\l\lb)^{-2}
[(\lb\g_{mnp}r)\Pi^p + (r\g_{mnp}r)\Gb^p] $$
$$
-{1\over{24}} \G^m\G^n\G^p [2(\l\lb)^{-3} (\lb\p\t)(\lb\g_{mnp}r) - 
(\l\lb)^{-2}(\lb\g_{mnp}\p\lb)]$$
where the constraint of \Gc\ has been used to substitute $\Gb^m$ for
$\half(\l\lb)^{-1} (\lb\g^m d) -{1\over 8} (\l\lb)^{-2} (\lb\g^{mnp}r) N_{np}$. 

One can also compute the gauge-invariant version of the stress tensor
and U(1) current of \purestress\ and \nonminu\
which are
\eqn\Tg{T= e^R ~ (-\half \p x^m \p x_m - p_\a \p\t^\a + w_\a \p\l^\a 
- s^\a \p r_\a + \bar w^\a \p\lb_\a) ~ e^{-R}}
$$=
-\half \p x^m \p x_m - p_\a \p\t^\a + w_\a \p\l^\a -
s^\a \p r_\a + \bar w^\a \p\lb_\a - \Gb^m \p\G_m$$
and
\eqn\Jg{ J= e^R~ (w_\a\l^\a + r_\a s^\a)~ e^{-R} 
= 
w_\a\l^\a + r_\a s^\a + \G_m \Gb^m.}

The operators of \Tg, \bg, \BRSTg\ and \Jg\ form a set of twisted
N=2 superconformal generators which preserve the first-class
constraint of 
\Gc. The resulting N=2 superconformal field theory will
be related to a dynamical twisting of the RNS formalism where
the RNS fermionic vector variable $\psi^m$ is defined as
\eqn\psidef{\psi^m = \Gb^m + \half (\l\lb)^{-1} \G_n (\l\g^m\g^n\lb).}
Note that $\psi^m$ satisfies the usual OPE $\psi^m(y)\psi^n(z) 
\to (y-z)^{-1} \eta^{mn}$ and commutes with the constraint 
$\Gb^m(\g_m\lb)^\a=0$. Since this constraint eliminates half of the
$\Gb^m$ variables and can be used to gauge-fix half of the $\G_m$
variables, the remaining 10 variables of $\Gb^m$ and $\G_m$
can be expressed in terms of $\psi^m$.

Although $w_\a$ and $\bar w^\a$ have singular OPE's with $\psi^m$,
one can define variables $w'_\a$ and $\bar w'^\a$ which have no singular
OPE's with $\psi^m$ as
\eqn\wp{w_\a = w'_\a -{1\over 4} \psi_m\psi_n [ (\l\lb)^{-1} (\g^{mn}\lb)_\a
- \lb_\a (\l\lb)^{-2} (\l\g^{mn}\lb) ],}
$$\bar w^\a -\half \Gb^m \G^n (\l\lb)^{-1} (\g_m\g_n\l)^\a = 
\bar w'^\a -{1\over 4} 
\psi_m\psi_n [ (\l\lb)^{-1} (\g^{mn}\l)^\a
- \l^\a (\l\lb)^{-2} (\lb\g^{mn}\l) ].$$
Note that $\bar w^\a$ always appears in the combination 
$\bar w^\a -\half \Gb^m \G^n (\l\lb)^{-1} (\g_m\g_n\l)^\a$
since it is this combination which commutes with the constraint $\Gb^m
(\g_m\lb)^\a =0$.

When expressed in terms of $\psi^m$, $w'_\a$ and $\bar w'^\a$, the 
twisted N=2 generators of 
\Tg, \bg, \BRSTg\ and \Jg\ take the form
\eqn\twistprime{
T = 
-\half \p x^m \p x_m - p_\a \p\t^\a + w'_\a \p\l^\a -
s^\a \p r_\a + \bar w'^\a \p\lb_\a}
$$ -\half \psi^m \p\psi_m
-{1\over 4}\p[ (\l\lb)^{-1}
(\l\g_m\g_n\lb) \psi^m \psi^n],$$

$$G^- = \half (\l\lb)^{-1} 
(\l\g_m\g_n\lb) \psi^m \Pi^n +
s^\a\p\lb_\a + w'_\a \p\t^\a -\half (\l\lb)^{-1} (w'\g^m\lb)
(\l\g_m\p\t)$$
$$+ {1\over 4}
\psi_m\psi_n (\l\lb)^{-1} [(\lb\g^{mn}\p\t) 
+(\l\lb)^{-1} (\lb\p\t)(\l\g^{mn}\lb)
+ (r\g^{mn}\l)+
(\l\lb)^{-1} (r\l))(\l\g^{mn}\lb)], $$

$$ G^+ = -\half (\l\lb)^{-1}
(\l\g_m\g_n\lb) \psi^n \Pi^m +
\l^\a d_\a - \bar w'^\a r_\a $$
$$+{1\over 4} \psi_m\psi_n (\l\lb)^{-1} [(\lb\g^{mn}\p\t) 
+(\l\lb)^{-1} (\lb\p\t)(\l\g^{mn}\lb)
+ (r\g^{mn}\l)+
(\l\lb)^{-1} (r\l))(\l\g^{mn}\lb)], $$
$$+ G^- [{1\over{24}}(\l\lb)^{-2} (\lb\g_{mnp}r) \psi^m\psi^n\psi^p ],$$

$$J =- \half(\l\lb)^{-1}(\l\g_{mn}\lb)\psi^m\psi^n + w'_\a \l^\a
+ r_\a s^\a,$$
where 
$G^- [{1\over{24}}(\l\lb)^{-2} (\lb\g_{mnp}r) \psi^m\psi^n\psi^p ]$
denotes the single pole in the OPE of $G^-$ with
${1\over{24}}(\l\lb)^{-2} (\lb\g_{mnp}r) \psi^m\psi^n\psi^p$ and is equal
to the last two lines of \BRSTg.

Except for the extra term
$G^- [{1\over{24}}(\l\lb)^{-2} (\lb\g_{mnp}r) \psi^m\psi^n\psi^p ]$
in $G^+$, the generators of \twistprime\ have a very symmetric form.
This asymmetry in $G^+$ and $G^-$ can be removed by performing the
similarity transformation ${\cal O} \to e^{R} ~{\cal O}~ e^{-R}$ on all
operators where 
\eqn\simtwo{R= -{1\over{24}}
\int (\l\lb)^{-2} (\lb\g_{mnp} r)\psi^m\psi^n\psi^p.}
This similarity transformation leaves $G^+$ of \twistprime\
invariant but transforms $T$, $G^-$ and $J$ as
\eqn\trans{ T \to T + {1\over{24}}
\p((\l\lb)^{-2} (\lb\g_{mnp} r)\psi^m\psi^n\psi^p),}
$$G^- \to G^- +  G^-[{1\over{24}}(\l\lb)^{-2} (\lb\g_{mnp}r) 
\psi^m\psi^n\psi^p ],$$
$$J \to J + {1\over{12}}
(\l\lb)^{-2} (\lb\g_{mnp} r)\psi^m\psi^n\psi^p.$$
It also transforms the constraint of 
\Gc\ into the constraint
\eqn\Gcnew{
\half(\l\lb)^{-1}(\l\g^n\g^m\lb)\psi_n = 
\half(\l\lb)^{-1} (\lb\g^m d) -{1\over 8} (\l\lb)^{-2} (\lb\g^{mnp}r) N'_{np} }
where $N'_{np} = \half w' \g_{np} \l$.

After performing the similarity transformation of \simtwo, the twisted
N=2 generators preserve the constraint of \Gcnew\
and take the symmetrical form
\eqn\twistprimenew{
T = 
-\half \p x^m \p x_m - \half \psi^m \p \psi_m 
- p_\a \p\t^\a + 
\half (w'_\a \p\l^\a -\l^\a \p w'_\a) }
$$
-\half (s^\a \p r_\a + r_\a \p s^\a) + \bar w'^\a \p\lb_\a + \half \p J,$$

$$-G^+ + G^- = \psi_m \Pi^m - \l^\a d_\a + \bar w'^\a r_\a +
s^\a\p\lb_\a + w'_\a \p\t^\a -\half (\l\lb)^{-1} (w'\g^m\lb)
(\l\g_m\p\t),$$

$$J =-\half (\l\lb)^{-1}(\l\g_{mn}\lb)\psi^m\psi^n + 
{1\over{12}}(\l\lb)^{-2} (\lb\g_{mnp} r)\psi^m\psi^n\psi^p +
w'_\a \l^\a
+ r_\a s^\a,$$

$$G^+ + G^- = [-G^+ + G^-, J]$$
$$  = \psi_m\Pi_n (\l\lb)^{-1} (\l\g^{mn}\lb)
+ \l^\a d_\a - \bar w'^\a r_\a +
s^\a\p\lb_\a + w'_\a \p\t^\a -\half (\l\lb)^{-1} (w'\g^m\lb) $$
$$ +\half \psi_m\psi_n (\l\lb)^{-1} [(\lb\g^{mn}\p\t) 
+(\l\lb)^{-1} (\lb\p\t)(\l\g^{mn}\lb)
+ (r\g^{mn}\l)+
(\l\lb)^{-1} (r\l))(\l\g^{mn}\lb)] $$
$$ + {1\over 4}\psi^m\psi^n [(\l\lb)^{-2}
(\lb\g_{mnp}r)\Pi^p +\half (\l\lb)^{-3}(r\g_{mnp}r)(\lb\g^p\g^q\l)\psi_q]$$
$$ +{1\over{12}}\psi^m\psi^n\psi^p [-2(\l\lb)^{-3}
(\lb\p\t)(\lb\g_{mnp}r) + (\l\lb)^{-2} 
(\lb\g_{mnp}\p\lb) ],$$
where the last two lines in $G^+ + G^-$ is
$G^- [{1\over{12}}(\l\lb)^{-2} (\lb\g_{mnp}r) \psi^m\psi^n\psi^p ]$.
These N=2 generators of \twistprimenew\
will now be related to a dynamically twisted version of the RNS
formalism.

\newsec{Dynamical Twisting of the RNS Formalism}

In this section, the RNS
formalism will be ``dynamically twisted'' to an N=2 superconformal
field theory by introducing bosonic pure spinor variables
$\l^\a$ and $\lb_\a$ and their fermionic worldsheet superpartners.
The corresponding twisted N=2 superconformal generators will then be
related to the twisted N=2 generators of \twistprimenew\
in the pure spinor formalism.

Twisting the N=1 RNS superconformal generators
\eqn\rns{T = -\half \p x^m \p x_m -\half \psi^m \p \psi_m, \quad
G = \psi^m \p x_m}
into N=2 superconformal generators usually involves choosing a U(5)
subgroup of the Wick-rotated
$SO(10)$ Lorentz group and splitting the ten $x^m$ and $\psi^m$ variables
into five complex pairs $(x^a, \bar x^{\bar a})$ 
and $(\psi^a, \bar\psi^{\bar a})$ for $a=1$ to 5. One then defines the twisted
N=2 superconformal generators as
\eqn\twistrns{T_{RNS} = 
- \p x^a \p x^{\bar a} - \bar\psi^{\bar a} \p\psi^a,}
$$
G^-_{RNS} = \bar\psi^{\bar a} \p x^a,  \quad
G^+_{RNS} = -\psi^a \p\bar x^{\bar a}, $$
$$J_{RNS} = -\bar\psi^{\bar a} \psi^a,$$
which satisfy the OPE
$G^+(y) G^-(z) \to (y-z)^{-2} J(z) + (y-z)^{-1} T(z)$.

To dynamically twist, one instead introduces pure spinor worldsheet variables
$\l^\a$ and $\lb_\a$ satisfying 
\eqn\purec{\l\g^m\l=0, \quad \lb\g^m\lb=0,}
whose
projective components parameterize the coset $SO(10)/U(5)$.
The N=2 superconformal generators of \twistrns\ can then be written in
a Lorentz-covariant manner as
\eqn\twistrns{T_{RNS} = -\half \p x^m \p x_m -\half \psi^m \p\psi_m 
-{1\over 4} \p[(\l\lb)^{-1}(\l\g^m \g^n\lb)\psi_m \psi_n], }
$$G^-_{RNS} = \half(\l\lb)^{-1} (\l\g^m\g^n\lb)\psi_m \p x_n,
\quad G^+_{RNS} = -\half(\l\lb)^{-1} (\l\g^n\g^m\lb)\psi_m \p x_n, $$
$$ J_{RNS} =-\half (\l\lb)^{-1} (\l\g^m\g^n\lb)\psi_m \psi_n.$$

The next step is to introduce the fermionic worldsheet superpartners of the pure
spinor variables $(\l^\a, \lb_\a)$ and their conjugate momenta 
$(w'_\a, \bar w'^\a)$. The fermionic superpartners of
$\l^\a$ and $w'_\a$ will be denoted $\tilde \t^\a$ and $\tilde p_\a$, and 
the fermionic superpartners of
$\lb_\a$ and $\bar w'^\a$ will be denoted $r_\a$ and $s^\a$. 
They are 
constrained to satisfy 
\eqn\supercons{\l\g^m\p\tt =0, \quad \lb \g^m r=0,} 
which will be 
the worldsheet supersymmetry transformation of the pure spinor constraints of 
\purec.
Because of the constraint $\l\g^m\p\tt=0$, $\tt^\a$ is a constrained version
of $\t^\a$ which only contains eleven independent non-zero modes.
The corresponding twisted N=2 superconformal generators for
these pure spinor multiplets are defined as
\eqn\twistpure{T_{pure} = w'_\a \p\l^\a - \tilde p_\a \p \tt^\a + 
\bar w'^\a \p\lb_\a - s^\a \p r_\a, }
$$G^-_{pure} =  w'_\a\p\tt^\a + s^\a \p\lb_\a, \quad
G^+_{pure} = \l^\a \tilde p_\a - \bar w'_\a r^\a,$$
$$ J_{pure} = w'_\a \l^\a + r^\a s_\a,$$
which preserve the pure spinor constraints of \purec\ and \supercons.

Finally, one adds the N=2 superconformal generators of \twistrns\ and
\twistpure\ in a manner that preserves the N=2 algebra. This can be done by
defining $T$, $J$ and $-G^+ + G^-$ as the sum
\eqn\sum{T = T_{RNS} + T_{pure}, \quad J = J_{RNS} + J_{pure},}
$$-G^+ + G^- = (-G^+ + G^-)_{RNS} + (-G^+ + G^-)_{pure}, $$
and then defining $G^++G^-$ using the commutator algebra
$$G^+ + G^- = [- G^+ + G^-, J].$$
Since $G^+_{pure}$ and $G^-_{pure}$ do not commute with $J_{RNS}$, $G^++G^-$
is not the sum of $(G^++G^-)_{RNS}$ and $(G^++G^-)_{pure}$.

The resulting N=2 superconformal generators for the dynamically
twisted RNS formalism are
\eqn\result{T = -
\half \p x^m \p x_m - 
\half \psi^m \p\psi_m -
\tilde p_\a \p\tt^\a + \half (w'_\a \p\l^\a -\l^\a\p w'_\a) }
$$-
\half (s^\a \p r_\a +r_\a \p s^\a) + \bar w'^\a \p\lb_\a + \half \p J,$$

$$-G^+ + G^- = \psi^m \p x_m - \l^\a \tilde p_\a + \bar w'^\a r_\a +
s^\a\p\lb_\a + w'_\a \p\tt^\a,$$

$$J =-\half (\l\lb)^{-1}(\l\g_{mn}\lb)\psi^m\psi^n + w'_\a \l^\a
+ r_\a s^\a,$$

$$G^+ + G^- = [-G^+ + G^-, J]$$
$$= \psi_m \p x_n (\l\lb)^{-1} (\l\g^{mn}\lb) 
+ \l^\a \tilde p_\a - \bar w'^\a r_\a +
s^\a\p\lb_\a + w'_\a \p\tt^\a$$
$$+\half
\psi_m\psi_n (\l\lb)^{-1} [(\lb\g^{mn}\p\tt) 
+(\l\lb)^{-1} (\lb\p\tt)(\l\g^{mn}\lb)
+ (r\g^{mn}\l)+
(\l\lb)^{-1} (r\l))(\l\g^{mn}\lb)]. $$

The N=2 superconformal generators of \result\ are obviously closely
related to the N=2 generators of \twistprimenew\ in the
pure spinor formalism, but there
are three important differences. Firstly, the generators of \result\
are not manifestly spacetime supersymmetric since they involve $\p x^m$
and $\tilde p_\a$ instead of $\Pi^m$ and $d_\a$. Secondly, the 
U(1) generator $J$ of \result\ does not include the term
${1\over{12}}(\l\lb)^{-2}(\lb\g_{mnp}r)\psi^m\psi^n\psi^p$. And thirdly, the
$\tt^\a$ variable in \result\ is constrained to satisfy $\l\g^m\p\tt=0$.

The first difference is easily removed by performing the similarity
transformation ${\cal O} \to e^{R} ~{\cal O}~ e^{-R}$ on all operators
in \result\ where
\eqn\simthree{ R =\half \int (\l\g^m\tt)\psi_m.} 
This similarity transformation does not affect $T$ or $J$ of \result\ but
transforms $-G^+ + G^-$ into the manifestly spacetime supersymmetric expression
\eqn\sups{-G^+ + G^- = \psi^m \tilde\Pi_m - 
\l^\a \tilde d_\a + \bar w'^\a r_\a +
s^\a\p\lb_\a + w'_\a \p\tt^\a}
where $\tilde\Pi^m = \p x^m + \half(\tt\g^m\p\tt)$ and
$\tilde d_\a = \tilde p_\a -\half (\p x^m +{1\over 4}
 (\tt\g_m\p\tt))(\g_m\tt)_\a$,
and transforms the $\psi_m\p x_n (\l\lb)^{-1} (\l\g^{mn}\lb)$ term in $G^++G^-$
into
$\psi_m\tilde\Pi_n (\l\lb)^{-1} (\l\g^{mn}\lb)$.

The second difference in the generators can be removed by modifying the
definition of dynamical twisting in \twistrns\
so that the appropriate term is added
to $J$. The generator
$-G^+ + G^- = (-G_+ +G^-)_{RNS} + (-G^+ + G^-)_{pure}$ and the untwisted
stress tensor
$T-\half \p J = 
(T-\half \p J)_{RNS}+
(T-\half \p J)_{pure}$ of \result\
will be left unchanged. But $J$ will be modified so that
after performing the similarity transformation of \simthree, the new $J$
includes the term
${1\over{12}}(\l\lb)^{-2}(\lb\g_{mnp}r)\psi^m\psi^n\psi^p$. 
And to preserve the N=2 algebra, $G^+ + G^-$ will be defined as
the commutator $[-G^+ + G^-, J]$ using the new
$J$.

Since
$e^{-R}~\psi^m ~e^{R} = \psi^m - \half(\l\g^m\tt)$,
this means one should modify $J$ in \result\ to
\eqn\jrnsnew{ J =-\half (\l\lb)^{-1} (\l\g^{mn}\lb)\psi_m\psi_n + w'_\a \l^\a
+r_\a s^\a}
$$ +{1\over{12}}
(\l\lb)^{-2} (\lb\g^{mnp}r) 
(\psi_m -\half (\l\g_m\tt))
(\psi_n -\half (\l\g_n\tt))
(\psi_p -\half (\l\g_p\tt)).$$
Although this modification of $J$
looks unnatural, it has the important
consequence of breaking the abelian shift symmetry $\tt^\a \to
\tt^\a + c^\a$
where $c^\a$ is any constant. This shift symmetry leaves invariant
the generators of \result, but has no corresponding 
symmetry in the pure
spinor formalism and should not be a physical symmetry.

After modifying $J$ in this manner and performing the similarity
transformation of \simthree, the generators of \result\ coincide with
the generators of \twistprimenew\ except for the restriction that
$\l\g^m\p\tt=0$. This final difference between the generators can be
removed by interpreting $\l\g^m\p\tt=0$ as a partial
gauge-fixing condition for
the symmetry generated by the first-class constraint of \Gcnew.
After relaxing the restriction $\l\g^m\p\tt=0$ and adding
the term $-\half(\l\lb)^{-1}(w'\g^m\lb)(\l\g^m\p\t)$ to $G^-$, 
the generators of \result\
coincide with those of \twistprimenew\ and therefore preserve
the constraint of \Gcnew.

Since the generators preserve \Gcnew, 
it is consistent to interpret \result\ as a partially gauge-fixed version
of \twistprimenew\ where the symmetry generated by \Gcnew\ is used to
gauge-fix $\l\g^m\p\t=0$.
On the other hand, the original N=2 generators of \purestress\ --
\nonminu\
of the pure spinor
formalism can be interpreted as a gauge-fixed version of \twistprimenew\
where the gauge-fixing condition is $(\l\g^m\g^n\lb)\psi_n=0$.
This is easy to see since 
$(\l\g^m\g^n\lb)\psi_n=0$ implies that $R=0$ in the similarity transformations
of \BRSTg, \Tg\ and \Jg.

\newsec{Summary}

In section 2, the $b$ ghost of the pure spinor formalism was simplified
by introducing the fermionic vector variable $\bar\G^m$ of \Gc.
After expressing $\bar\G^m$ in terms of the RNS variable $\psi^m$ using
\psidef, the $b$ ghost and BRST current form a symmetric set of twisted
N=2 generators \twistprimenew\ which preserve the constraint of \Gcnew.

In section 3, the corresponding N=2 superconformal field theory was
interpreted as a dynamically twisted version of the RNS formalism in which
the pure spinors $\l^\a$ and $\lb_\a$ parameterize the $SO(10)/U(5)$ choices
of twisting. The dynamically twisted RNS generators are obtained from
\twistprimenew\ using the constraint of \Gcnew\ to gauge-fix $\l\g^m\p\t=0$.
And the twisted N=2 
generators of the original pure spinor formalism are obtained
from \twistprimenew\ 
using the constraint of \Gcnew\ to gauge-fix $(\l\g^m\g^n\lb)\psi_n=0$.

\vskip 10pt
{\bf Acknowledgements:}
I would like to thank Renann Jusinskas, 
Nikita Nekrasov and Edward Witten for useful discussions,
and 
CNPq grant 300256/94-9
and FAPESP grants 09/50639-2 and 11/11973-4 for partial financial support.

\listrefs
\end